\documentclass[twocolumn,
superscriptaddress,
prl,
showpacs,
preprintnumbers,
amsmath,amssymb]{revtex4}

\usepackage{graphicx,epsf,epsfig,ulem}
\usepackage{epsfig}
\usepackage{bm}
\usepackage{subfigure}
\usepackage{color}

\begin{document}

\title{Stark tuning of the charge states of a two-donor molecule in silicon}

\author{Rajib Rahman}
\affiliation{Network for Computational Nanotechnology, Purdue University, West Lafayette, IN 47907, USA}

\author{Seung H. Park}
\affiliation{Network for Computational Nanotechnology, Purdue University, West Lafayette, IN 47907, USA}


\author{Gerhard Klimeck}
\affiliation{Network for Computational Nanotechnology, Purdue University, West Lafayette, IN 47907, USA} 
\affiliation{Jet Propulsion Laboratory, California Institute of Technology, Pasadena, CA 91109, USA}

\author{Lloyd C. L. Hollenberg}
\affiliation{Center for Quantum Computer Technology, School of Physics, University of Melbourne, VIC 3010, Australia}

\date{\today} 

\begin{abstract} 
Gate control of phosphorus donor based charge qubits in Si is investigated using a tight-binding approach. Excited molecular states of $\textrm{P}_2+$ are found to impose limits on the allowed donor separations and operating gate voltages. The effects of surface (S) and barrier (B) gates are analyzed in various voltage regimes with respect to the quantum confined states of the whole device. Effects such as interface ionization, saturation of the tunnel coupling, sensitivity to donor and gate placement are also studied. It is found that realistic gate control is smooth for any donor separation, although at certain donor orientations the S and B gates may get switched in functionality. This paper outlines and analyzes the various issues that are of importance in practical control of such donor molecular systems. 
\end{abstract} 


\maketitle

\section {I. Introduction}

Single donor systems in silicon have been the subject of much research in recent years from both theorists and experimentalists alike. While a number of experiments have probed into the physics of single donors \cite{Brandt.nature.2006, Brandt.prl.2006, Rogge.NaturePhysics.2008, Sellier.prl.2006, Bradbury.prl.2006}, many others have concentrated on precision control and fabrication of qubits based on individual donors \cite{Schofield.prl.2003, Jamieson.apl.2005}. Apart from the initially predominant effective mass based approaches for donor modeling \cite{Koiller.prl.2002, Friesen.prl.2005}, a number of other techniques have been successfully applied to study realistic devices and interpret experimental results in detail \cite{Wellard.prb.2005, Martins.prb.2004, Rahman.prl.2007}.

Although the initial qubit proposals utilized the spin of the donor electron or nucleus to encode qubits \cite{Kane.nature.1998, Vrijen.pra.2000, Skinner.prl.2003}, one can also envisage fabricating and controlling the charge degrees of freedom of a simple donor molecule: a singly ionized double donor system \cite{Hollenberg.prb.2004}. Steps along this direction have been achieved with charge state transfer between two P donors in silicon reported \cite{Andresen.Nanoletter.2007}, and a single electron in a donor-interface double well forming a new hybrid molecular system \cite{Rogge.NaturePhysics.2008}. Despite possessing the disadvantages of shorter decoherence times, such a charge qubit system is easier to probe experimentally and may help to demonstrate the feasibility of theoretical quantum computation concepts. The donor based charge qubit is also more amenable to measurements as the donor electron can be localized to particular impurities with relative ease by means of appropriate gate placement and voltage pulses. Microwave driven experiments have already been proposed to investigate the parameter space and physical operation of such devices \cite{Wellard.prb.2006}. Precision placement of a few donors has also been achieved recently, highlighting the urgency for a detailed theoretical study of a full charge qubit device before its physical realization is achieved \cite{Andresen.Nanoletter.2007}. Charge qubits also form the building blocks of the coherent charge transfer mechanism proposed in Ref \cite{Greentree.prb.2004}. In such a scheme, quantum in formation encoded in the spin or in the charge of a donor electron can be transported coherently through a chain of ionized donors by an adiabatic pathway realized through voltage pulses applied to electrical gates. Termed as Coherent Tunneling Adiabatic Passage (CTAP), this mechanism may provide a robust way to transfer information in a circuit of donor spin qubits \cite{Rahman.prb.2009, Hollenberg.prb.2006}. 

An important parameter of interest in a $\textrm{P}_2+$ molecule is the tunnel coupling between the two impurity quantum dots defined by the coulomb potential of each donor nucleus. This coupling is expressed as the difference between the two lowest eigen states of the molecular system. A previous work \cite{Xuedong.prb.2005} had utilized Kohn-Luttinger \cite{Kohn.physrev.1955} type effective mass based variational envelope functions modulated by Bloch states of the six conduction band valleys of Si to show that the tunnel coupling suffers from the same sensitivity to relative donor placements as the inter-donor exchange coupling for spin qubits. In Ref \cite{Koiller.prb.2006}, the same approach was used to show that the tunnel coupling can be controlled smoothly by constant electric fields for both homo and hetero polar donor species. As experiments are getting closer to fabricating an actual donor based charge qubit system \cite{Andresen.Nanoletter.2007}, it is important to develop a detailed theoretical model that analyzes the various aspects of qubit control and design space, and predicts accurate numbers for quantities likely to be measured by experiments. Our goal in this paper is to serve this very purpose by modeling the $P_2+$ molecule with a comprehensive numerical approach.  

We employ atomistic tight-binding theory with a model for P impurity in Si that accurately captures the basic single donor physics such as the valley-orbit split donor spectrum. By solving the full tight-binding Hamiltonian for realistic systems of about 3 million atoms, we are able to obtain any number of states of the $\textrm{P}_2+$ spectrum. Previous works \cite{Xuedong.prb.2005, Koiller.prb.2006} on donor charge qubits have ignored the effects of the excited manifold on device operations. However, in the presence of significant gate bias, these excited states can disrupt normal device operation by entering the manifold of states used in quantum computing operations. Furthermore, effects of realistic gate potential profiles on the tunnel coupling need to be investigated as opposed to the simplistic constant electric fields used in earlier works. A gate can also cause surface ionization of the bound electron at higher biases, thus limiting the control regime in practice. Presence of nearby interfaces can significantly distort the donor wave functions and affect charge qubit operations. Since the previous works concentrated on bulk systems, these effects have been neglected. 

In our tight-binding model coupled with electrostatic gate potentials from a commercial Poisson solver \cite{Ise.TCAD}, we can investigate all these effects under one framework with considerable accuracy. It is also important to go beyond the effective mass theory (EMT) assumptions of only the valley minima states contributing to the donor wave functions, and to consider a more comprehensive Bloch structure of Si as is done in tight-binding. In the case of inter-qubit exchange coupling for spin qubits, Wellard \cite{Wellard.prb.2003} showed that oscillations of the exchange coupling J(V) could be damped to some extent if an extended set of Bloch states is considered rather than the six valley minima states to expand the donor wave function. Wellard et al. \cite{Wellard.prb.2003} also calculated the angular dependence of J(V) for a fixed radial separation of donors, and showed that a gate bias was not able to alter J(V) significantly for certain angular separations between donors. It remains to be seen if the tunnel coupling also suffers from this controllability problem. 

This paper is organized into three parts. In Section I, the geometry of a P donor based charge qubit in Si is described, and the various control parameters studied here are outlined. In Section II, the details of the tight-binding method is described. Section III presents the results and discusses the controllability issues in detail.

\section {II. $\textrm{P}_2+$ donor device geometry}

\begin{figure}[htbp]
\center
\epsfxsize=2.8in
\epsfbox{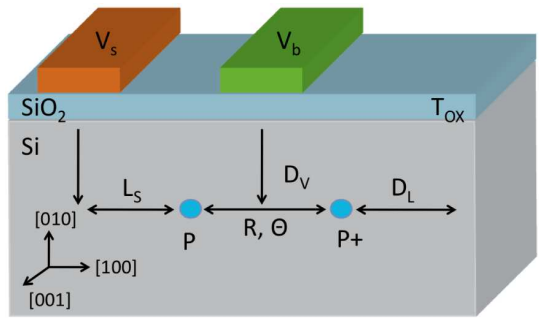}
\caption{The schematic of a donor based charge qubit device showing the various design parameters. The surface gate voltage $V_S$ detunes the charge states of the donor molecule, while the barrier gate voltage $V_B$ controls the tunneling barrier between the donors.} 
\end{figure} 

A schematic of a $\textrm{P}_2+$ donor molecular device in [001] grown Si is illustrated in Fig 1. The P donors are separated by a distance $R$ nm and an angle $\Theta$ measured from the [100] axis. The donors form a molecular system analogous to $\textrm{H}_2+$ except the Si band structure complicates the scenario. A barrier gate (B-gate) is placed midway between the impurities and controls the potential barrier between them. A surface gate (S-gate) is placed a distance $L_S$ (measured with respect to the center of the S-gate) away from the left impurity, and controls the detuning of the impurity states. the oxide thickness $T_{ox}$ is 5 nm, while the gate lengths are 10 nm. The impurities are buried at a depth $D_V$ below the oxide and a distance $D_L$ from the lateral interfaces. Typical devices have dimensions 50 nm $\times$ 40 nm $\times$ 30 nm. While a larger device volume would be ideal, the compute times increase much more making the problem intractable. This device volume, however, captures most of the Physics of the double donor molecule.

First, we study the molecular spectrum of the system without any gate voltage taking into account the complicated band structure of Si. Then we investigate how this spectrum, and in particular the tunnel coupling, can be controlled by surface and barrier gates. Effects of design parameters such as $L_S$, $D_V$, $R$, $\Theta$, $V_S$, and $V_B$ are also explored. In particular, the energy gaps $\Delta_{12}$ and $\Delta_{23}$ are studied. The energy gap $\Delta_{12}$ represents the energy difference between the 1st excited state ($\textrm{S}_2$) and the ground state ($\textrm{S}_1$), while $\Delta_{23}$ represents the energy difference between the 2nd excited state($\textrm{S}_3$) and the 1st. The device schematic and notations of Fig 1 is used throughout this paper. The molecular states are labeled as $\textrm{S}_1$, $\textrm{S}_2$, $\textrm{S}_3$, and so on, from more strongly bound to less ($\textrm{S}_1$ being the ground state).

\section {III. Tight binding solution of the solid-state donor molecular system}

The approach used here is the semi-empirical tight-binding (TB) method \cite{Slater.physrev.1954} with the 20 band $sp^3d^5s*$ nearest-neighbor model. In this method, the Hamiltonian is expressed in real space with a basis of localized atomic orbitals. The 20 band TB parameters for Si are optimized by genetic algorithm \cite{Boykin.prb.2004, Klimeck.cmes.2002} to reproduce the bulk band structure of the host. Once a set of such parameters are found, it can be used for atomistic modeling of a generic device made of the host. A single P impurity is represented by a coulomb potential screened by the dielectric constant of Si ($\epsilon_{Si}$) and subjected to a cut-off potential $U_0$ at the impurity site. This core correcting potential $U_0$ is adjusted to reproduce the single donor ground state binding energy. The total TB Hamiltonian is of the form,

 \begin{eqnarray} 
 \label{eq:hamiltonian} 									
H &=& H_0+eV_{D1}+eV_{D2}+\frac{e^2}{4\pi\epsilon_{Si} |R_1-R_2|} \nonumber \\
&& {} +eV_G(V_S, V_B) 
\end{eqnarray}

\noindent  
where the first term is the crystal Hamiltonian of Si, the second and third terms the potential energy due to the two donor nulcei, the 4th term the nuclear repulsion of the two positively charged impurity cores, and the last term the potential energy due to the surface and barrier gates obtained from a commercial Poisson Solver \cite{Ise.TCAD}. The single impurity potential is expressed as, 

\begin{eqnarray} 
V_{Di}(r)& = &-\frac{e}{4\pi\epsilon_{Si} |r-R_i|}, \qquad r \neq R_i  \\ 
V_{Di}(r)& = &-U_{0}, \qquad \qquad \qquad r = R_i
\end{eqnarray}
where $R_i$ is the location of the $i$-$th$ impurity. Interfaces are treated by closed boundary conditions with a model of surface passivation of dangling bonds \cite{Lee.prb.2004}. The full atomistic Hamiltonian of about 3 million Si atoms was solved by a parallel Lanczos / Block Lanczos algorithm to obtain the eigenvalues and wavefunctions in the desired energy range \cite{Klimeck.cmes.2002}. Each data point in this work required about 6 hours on 48 processors \cite{nanohub.note}. 

The tight-binding method used here is under the hood of the the Nanoelectronic Modeling Tool (NEMO-3D) \cite{Klimeck.cmes.2002, Klimeck.ted.2007}, and had been successfully used to verify Stark shift of the hyperfine coupling between a donor and its nucleus \cite{Rahman.prl.2007} in good agreement with experiments \cite{Bradbury.prl.2006} and with momentum space methods \cite{Wellard.prb.2005}. The same method was also applied to investigate the orbital Stark shift of a donor-interface well system, and was verified with single donor transport experiments in FinFETs \cite{Rogge.NaturePhysics.2008}. 

\section {IV. Results and Discussions}

\subsection{A. The molecular spectrum of $\textrm{P}_2+$ at zero gate bias}

The lowest 6 states of a single Group V donor in Si are of 1s type due to the six-fold degenerate conduction band minima of Si. The sharp confining potential in the vicinity of the donor nucleus causes coupling between the valleys, a phenomena termed as valley-orbit (VO) or chemical splitting. The net effect of the VO interaction is a splitting of the 6 1s states into a singlet ($\textrm{A}_1$), triplet ($\textrm{T}_2$) and a doublet ($\textrm{E}_1$) orbital manifolds. The strength of this VO interaction varies from one donor species to another, and is caused by a number of microscopic properties such as variation of the dielectric constant of Si from its bulk value, local strain originating from bonds between the donor and the Si atoms, and so on. For an isolated P donor in Si, the ($\textrm{A}_1$), ($\textrm{T}_2$) and ($\textrm{E}_1$) states are bound at -45.6, -33.9, and -32.6 meV respectively. When two hydrogenic donors are placed near each other, each pair of corresponding 1s states gives rise to a bonding (symmetric) and an anti-bonding (anti-symmetric) state.  

\begin{figure}[htbp]
\center\epsfxsize=3.4in\epsfbox{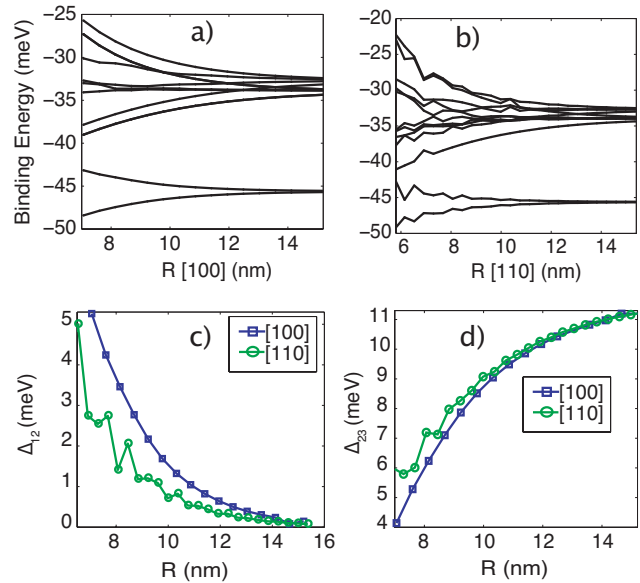}
\caption{Spectrum of the donor molecule showing the lowest 12 molecular orbital energies as a function of donor separation along a) [100], and b) [110]. c) and d) show the tunnel couplings between states 1 and 2 ($\Delta_{12}$), and 2 and 3 ($\Delta_{23}$) respectively.} 
\end{figure} 

Fig. 2 shows the variation of the binding energies of the 12 lowest molecular states of the $\textrm{P}_2+$ system as a function of impurity separation. The binding energies show smooth exponential decay with donor separation $R$ for impurities separated along [100] (Fig. 2a), but exhibit some oscillatory behavior for impurities separated along [110] (Fig 2b), consistent with EMT descriptions \cite{Xuedong.prb.2005}. For large enough $R$, the bonding and anti-bonding pairs become almost degenerate, and the binding energies reduce to the single P binding energies with double degeneracies. At small $R$, the bonding state arising from one of the $\textrm{T}_2$ states of each donor approaches the anti-bonding state arising from $\textrm{A}_1$ states, causing $\Delta_{23}$ to be comparable in magnitude to $\Delta_{12}$. This situation is undesirable in quantum computer architectures since the qubit Hilbert space, which usually consists of the two lowest eigenstates, needs to be well-isolated in energy from the rest of the manifold. A well-isolated qubit Hilbert space makes the qubit less prone to decoherence and errors during operation. In this paper, we will use $\Delta_{12} = \Delta_{23}$ as a threshold reference level for device operation. If $\Delta_{12} > \Delta_{23}$, the device is in an undesirable operation regime. From Fig 2a and 2b, it is observed that the donor separations need to be at least 7 nm for such coherent applications. 

\subsection{B. Surface gate control}

\begin{figure}[htbp]
\center\epsfxsize=3.4in\epsfbox{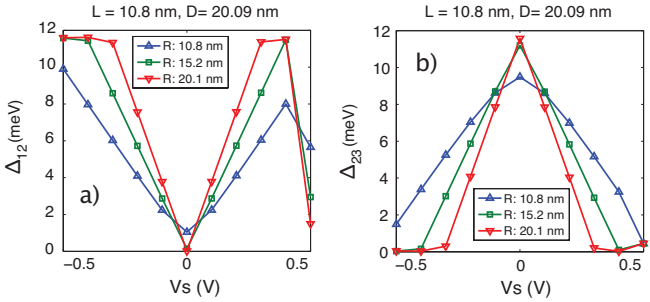}
\caption{Surface gate response of a) $\Delta_{12}$ and b) $\Delta_{23}$ for three different donor separations. The surface gate is located 10.8 nm ($L_S$) from the left donor, while the donors are buried 20.1 nm ($D_v$) below the oxide.} 
\end{figure} 

The function of a surface gate is to provide a potential difference between the two impurities, and hence to control the energy spacing of the eigen states of one impurity relative to the other. In the device geometry shown in Fig 1, a positive bias to the surface gate lowers the potential of the left impurity relative to the right impurity. The $\textrm{A}_1$ state of the left donor moves farther below that of the right donor, and this increases $\Delta_{12}$. The $\textrm{T}_2$ state of the left donor moves closer to the $\textrm{A}_1$ state of the right, which decreases $\Delta_{23}$. Fig 3a and 3b show the variation of $\Delta_{12}$ and $\Delta_{23}$ respectively with an S-gate voltage for three different donor separations along [100]. For larger impurity separations, the slopes of the $\Delta_{12}$ and $\Delta_{23}$ curves are steeper. If the impurity separations are larger, then the left impurity experiences a stronger S-gate potential than the right impurity. If the impurities are closer, then the same surface gate voltage provides less potential drop between the two impurities, and the surface gate response is weaker.   

To elucidate the different voltage regimes shown in Fig 3, it is helpful to look at the first three eigen states of the whole device (top 3 rows of Fig 4) along with the net electrostatic potential profile the device is operating under (bottom row of Fig 4). Each column in Fig 4 illustrates a snapshot of the donor molecule for a specific surface gate voltage with two donors separated by 15.2 nm along [100] (points marked by square data points in Fig 3).  

\begin{figure}[htbp]
\center\epsfxsize=3.4in\epsfbox{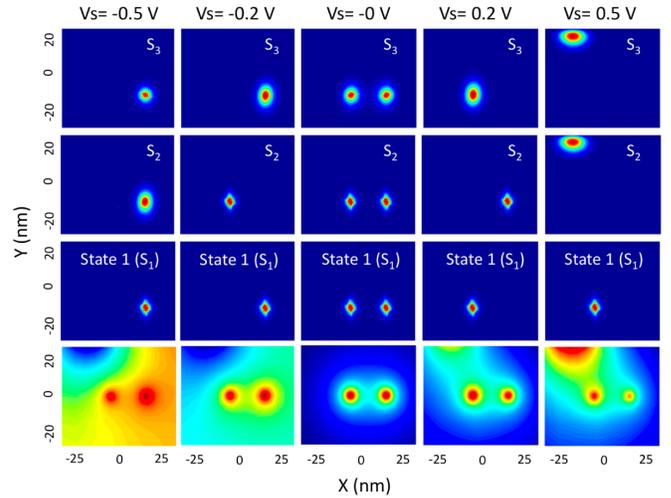}
\caption{The three lowest states of the donor molecule at various surface gate voltages. The bottom row shows the net electrostatic potential in the device, while the top 3 rows shows the first 3 wave function probability amplitudes. Each column represents the device at a specific voltage.} 
\end{figure} 

At $V_S = 0$ (col. 3), the two donor wells are aligned in energy, giving rise to bonding and anti-bonding pairs. The lowest two states ($\textrm{S}_1$ \& $\textrm{S}_2$) are formed from the symmetric and anti-symmetric superpositions of the $\textrm{A}_1$ state of each donor. $\textrm{S}_3$ is a bonding state arising from the $\textrm{T}_2$ states. 

As  $V_S$ is ramped up to 0.2 V (col. 4), the device is in the linear operation regime shown in Fig 3. The left impurity is lower in energy than the right, giving rise to a left localized ground state ($\textrm{A}_1$ of left P) and a right localized first excited state ($\textrm{A}_1$ of right P), while state 3 arises from the $\textrm{T}_2$ state of the left impurity. A plot of the total potential shows that the surface gate expands the potential contours of the left impurity relative to the right.  
 
If the S-gate voltage is reversed in polarity to -0.2 V (column 2), the effects described above are reversed between the two impurities, with a right localized state appearing as the ground state.

At $V_S = 0.5$ V (col. 5), a surface well is formed near the gate, and quantum states begin to appear in this well. In a finite sized nanostructure like this, whether the lowest states are at the surface or at the donor is determined by which well is lower in energy. At $V_S=0.5$ V, the surface well is almost as deep as the donor well. While the ground state of the system is still seen to occur in the left donor, the higher states occur in the surface well. A small increase in bias at this point moves the interface well deeper than the donor well, and produces ionization of the donor electron. For a donor close to the surface, the donor well can strongly couple to the interface well, and give rise to the prospect of adiabatic ionization \cite{ Kettle.prb.2003, Smit.prb.2004, Martins.prb.2004, Calderon.prl.2006, Rogge.NaturePhysics.2008}. Since the gate confined surface states are closely spaced in energy compared to the donor states, both $\Delta_{12}$ and $\Delta_{23}$ show a sharp decrease at the onset of this ionization process, as shown in Fig 3.  

At $V_S = -0.5$ V (col. 1), the interface well is raised in energy, and does not play a role. The gate voltage is high enough to push the $\textrm{T}_2$ state of the right donor below the $\textrm{A}_1$ of the left donor. When this happens, any subsequent change of the gate bias causes a small change in $\Delta_{12}$, and it flattens out, as shown in Fig 3a for the $R=$ 15.2 and 20.1 nm curves. This is expected as $\Delta_{12}$ is now equal to the splitting betweeb the $\textrm{A}_1$ and $\textrm{T}_2$ of the the same donor, and has a much weaker field response.  
The $\Delta_{23}$ curve captures the energy gap between the higher states of the same donor, and tends to flatten out also. The tunnel couplings therefore saturate in this voltage regime.

Since $\Delta_{12}$ is to be less than $\Delta_{23}$ for the donor device, the linear regimes about $V_S=0$ are important for coherent operations. The operational basis can be either the left and right localized states at modest bias (linear regime), or the bonding and anti-bonding states near zero bias \cite{Hollenberg.prb.2004}. The ionization regime at high positive bias can be important if the donor electron is to be shuttled from one donor to another by transport along the interface \cite{Skinner.prl.2003} or if a measurement at the surface is to be performed \cite{Calderon.prl.2006}.

\subsection{C. Barrier gate control}

\begin{figure}[htbp]
\center\epsfxsize=3.4in\epsfbox{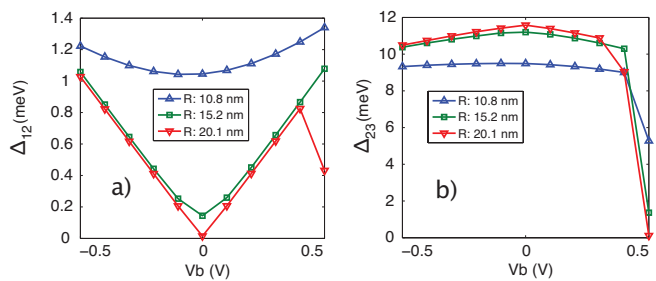}
\caption{Barrier gate response of a) $\Delta_{12}$ and b) $\Delta_{23}$ for three different donor separations with the donors 20.1 nm ($D_V$) below the oxide.} 
\end{figure} 

In contrast to a surface gate, a barrier gate subjects both donors to a similar potential. A positive barrier gate lowers the potential in the region between the two impurities, and increase hybridization of the left and right donor states. A negative barrier gate raises the potential of this region, and provides more localization of the single donor states to each impurity. $\Delta_{12}$ and $\Delta_{23}$ are affected less by B-Gates than S-Gates, as exhibited by the smaller slopes of the $\Delta_{12}$-$V_B$ curves of Fig 5 compared to those of Fig 3. The B-Gate also generates an interface well which eventually ionizes the donors at high enough gate bias, causing a sharp fall in $\Delta_{12}$ and $\Delta_{23}$.  
 
Fig 6 shows how the tunnel coupling is affected when both S-gate and B-gate biases are present simultaneously. In each curve, the B-gate is held fixed at some voltage, and the S-gate is varied over a range of -0.5 V to 0.5 V. It was shown in Fig 3 that at zero B-Gate bias, the $\Delta_{12}$-$V_S$ curve shows a minimum at $V_S = 0$. In the presence of a positive (negative) B-gate bias, this minimum shifts to a negative (positive) $V_S$ value. Since a positive B-gate lowers the potential in the barrier regions, and binds the electron more tightly between the donors, the symmetric-anti-symmetric gap increases. Although a negative S-gate bias causes detuning of the two impurities, it also lowers the potential in the barrier region, and thus compensates for the positive barrier gate. Hence, to obtain the same zero-field symmetric-anti-symmetric gap, a negative S-gate bias is needed, which explains the shift of the $\Delta_{12}$-$V_S$ curve towards a negative $V_S$ when $V_B$ is held positive. 

Since the gates are separated by 15 nm in this case, there is significant cross-talk between them. A positive S-gate therefore makes the surface well near B-gate deeper, and vice versa. As a result the ionization regime shifts to the left with increasing B-gate bias. 

Simultaneous S and B gate biases can be used to initialize the donor molecule system. At first, each donor has one electron bound to it. To form a $\textrm{P}_2+$ molecule, one electron has to removed. This can be done by holding $V_B$ negative so as to raise the barrier between the left and the right donors, while applying a simultaneous positive bias to the S-gate to ionize the left donor electron to the surface well.

\begin{figure}[htbp]
\center\epsfxsize=2in\epsfbox{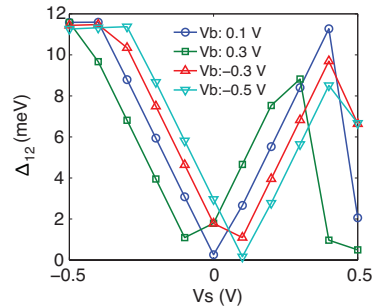}
\caption{Effect on $\Delta_{12}$ of applying biases to the surface and barrier gates simultaneously.} 
\end{figure} 

\subsection{D. Sensitivity to Donor Placement}

In this section, we investigate the sensitivity of the tunnel coupling to relative donor placements along different directions. In Fig 7, the radial separation between the donors is held fixed at 10.8 nm, while the angular separation is varied from $0$ to $45^0$ at a fixed impurity depth of 20.1 nm. $\Delta_{12}$ and $\Delta_{23}$ are plotted in Fig 7a and 7b respectively for three different gate configurations.   

\begin{figure}[htbp]
\center\epsfxsize=3.4in\epsfbox{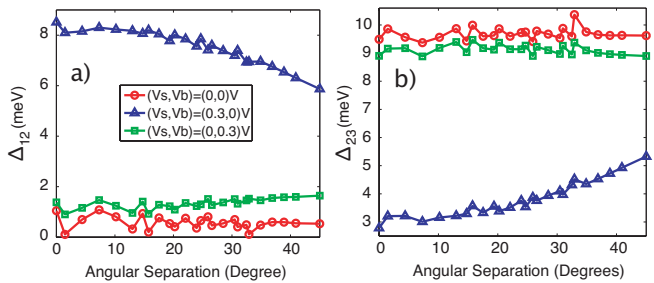}
\caption{a) $\Delta_{12}$ and b) $\Delta_{23}$ as a function of angular separation between the donors. The radial donor separation and the donor depths are held fixed at 10.86 nm and 20.1 nm respectively, while the angle is varied from $0$ to $45^0$ measured from the [100] direction.} 
\end{figure} 

Although the tunnel coupling shows some oscillatory behavior with angular separation, both a barrier and a surface gate voltage are each able to change the tunnel coupling significantly. This shows that there is enough gate controllability irrespective of how the donors are placed relative to each other. As a note of comparison, a similar study was done on the two-donor exchange coupling J(V) in Ref \cite{Wellard.prb.2003}. In contrast to the result obtained here, the J(V) curve exhibited some gate controllability issues. For certain angular separations between the donors, a gate voltage was not able to alter the magnitude of J(V) significantly. The absence of such controllability issues in this case is quite encouraging.

\begin{figure}[htbp]
\center\epsfxsize=3.4in\epsfbox{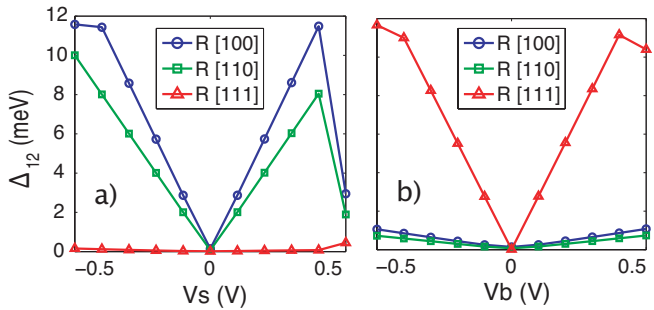}
\caption{a) Surface and b) barrier gate response of $\Delta_{12}$ for donors separated along three different directions. The device parameters: $R$=15.2 nm, $L_S$=10.86 nm, $D_V$=20.1 nm for R[100] and R[111].} 
\end{figure} 

Figures 3 and 5 demonstrated smooth gate control over donors separated along [100]. Fig 8 demonstrates that such smooth control also exists for donors separated along other directions. In addition, Fig 8 also demonstrates some geometry effects that need to be considered in practical devices. Fig 8a shows the effect of an S-gate on $\Delta_{12}$ for donors separated along three different directions, while Fig 8b shows the same for a B-Gate. 

The striking feature of the curves is the asymmetry in voltage control between the [111] separation and the other two directions.  If both impurities are not on the same xz plane (Fig 1), which happens for [111] separation, then a B-gate may subject one donor to a higher potential than the other. This is more pronounced if the donor depths below the gate are unequal by more than tens of lattice constants. In such a case, a B-gate can act as an S-gate, and the voltage response curve $\Delta_{12}$-$V_B$ can show a significantly larger response. The $\Delta_{12}$-$V_B$ curve of Fig 8b has a steeper slope for [111] donor separations compared to the [100] and [110] separations. 

Similarly, for [111] donor separations, an S-gate may be almost equidistant from the two donors, and can act as a B-gate, thus reducing the voltage response of $\Delta_{12}$ significantly. This is evident by the smaller slope of the [111] curve of Fig 8a compared to the other two curves. For the device structure considered here with donor separations along [111], the radial distances between the S-gate center and the left and right donors were 26.7 nm and 25.1 nm respectively. This shows that donor depth below the gate is an important feature in experimental design as surface gates can act as barrier gates and vice-versa for certain donor orientations.      

\subsection{E. Design issues: Gate placement and donor depth}

\begin{figure}[htbp]
\center\epsfxsize=3.4in\epsfbox{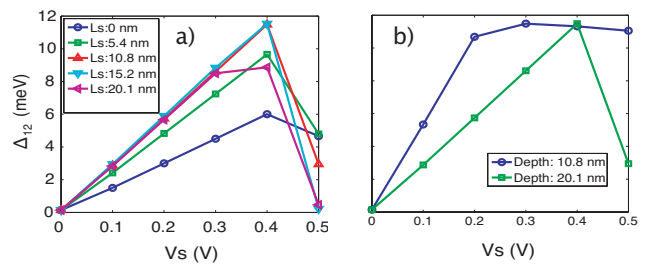}
\caption{Effect of surface gate placement on $\Delta_{12}$. b) S-gate response of $\Delta_{12}$ for different donor depths. The impurities are separated by 15.2 nm along [100].} 
\end{figure} 

Fig 9a shows the effect of varying the distance between the S-gate and the left impurity for two donors separated by 15.2 nm along [100]. The slope of the $\Delta_{12}$-$V_S$ curve is sensitive to $L_S$ because it determines the potential contours the two impurities reside on. are on similar potential contours. It is seen that the slopes of the curves are the steepest when $L_S$ is between 10.8 and 20.1 nm, suggesting that there may be an optimal gate distance from donor for design purposes. A small $L_S$ of 0 or 5.5 nm is likely to subject both impurities to a strong potential. As a result, the detuning is less responsive. 

In Fig 9b, we show how the S-gate response of the tunnel coupling is modified for two different donor depths below the oxide. For both donors at a shallower depth of 10.8 nm, the $\Delta_{12}$-$V_S$ curve is steeper because proximity of the donors to the gate lends more control. The curve is also seen to flatten off before reaching the ionization regime. This is because the stronger potential brings the $\textrm{T}_2$ state of left impurity below $\textrm{A}_1$ state of the right, and causes the tunnel coupling to become saturated before ionization is reached. Any further increase in $V_S$ after 0.2 V, simply captures the energy gap between the $\textrm{A}_1$ and $\textrm{T}_2$ states of the left impurity. In contrast, the curve for 20.1 nm depth reaches ionization directly after the linear regime.

\section{V. Conclusion}

Gate control of a $\textrm{P}_2+$ molecular system was investigated in detail from a tight-binding approach. It was found that excited states can place a limit on the range of operating voltages and donor separations. This arises from the necessity that the qubit Hilbert space needs to be sufficiently isolated from the other states of the system for robust coherent operation. Realistic TCAD \cite{Ise.TCAD} gate potentials were used to demonstrate that smooth controllability exists over the molecular states of the donors. The detuning of two donors by a surface gate in various voltage regimes was analyzed. It was also shown that the tunnel coupling is more sensitive to S-gate control than B-gate, with the exception of certain donor orientations for which the S-gate can act like a B-gate and vice versa. Simultaneous operation with both S and B gates was shown to shift the minimum energy gap to non-zero gate configurations. Practical control of such a system is also limited by interface ionization, and voltage regimes were established where this could take place. Sensitivity of the tunnel coupling to donor placement was also investigated. Although the tunnel coupling exhibits oscillatory behavior with donor separations in certain directions, it was found that a gate voltage was always able to magnify it significantly. This confirms that the tunnel coupling is free from the voltage controllability issue observed in J(V) \cite{Wellard.prb.2003}.  

The purpose of this work was to highlight the controllability and design issues that an experimental implementation and control of a $\textrm{P}_2+$ molecular system would involve. The focus here has been not only on the physical trends, but also on quantitative characterization of some quantities that are of interest to experimentalists. This $\textrm{P}_2+$ molecule forms the building blocks of structures that can utilize CTAP to transport information across qubits. Realization of CTAP in an actual donor array is likely to involve very precise gate control over quantum states of coupled donors. This work lays the foundation for further work on CTAP that involves precision gate control to realize an adiabatic path for population transfer across donor arrays.  
   
Acknowledgement: This work was supported by the Australian Research Council, the Australian Government, and the US National Security Agency (NSA), and the Army Research Office (ARO) under contract number W911NF-08-1-0527. Part of the development of NEMO-3D was done at JPL, Caltech under a contract with NASA. NCN/nanohub.org computational resources were used.

\noindent Electronic address: rrahman@purdue.edu \\
Electronic address: park43@purdue.edu


\vspace{-0.5cm}

\begin{thebibliography}{100}  

\bibitem{Brandt.nature.2006} A. R. Stegner, C. Boehme, H. Huebl, M. Stutzmann, K. Lips, and M. S. Brandt, Nature Physics {\bf{2}}, 835 (2006).

\bibitem{Brandt.prl.2006} Hans Huebl, A. R. Stegner, M. Stutzmann, M. S. Brandt, G. Vogg, F. Bensch, E. Rauls, and U. Gerstmann, Phys. Rev. Lett. {\bf{97}}, 166402 (2006). 

\bibitem{Rogge.NaturePhysics.2008} G. P. Lansbergen, R. Rahman, C. J. Wellard, I. Woo, J. Caro, N. Collaert, S. Biesemans, G. Klimeck, L. C. L. Hollenberg, and S. Rogge, Nature Physics {\bf{4}}, 656 (2008).

\bibitem{Sellier.prl.2006} H. Sellier, G. P. Lansbergen, J. Caro, S. Rogge, N. Collaert, I. Ferain, M. Jurczak, and S. Biesemans, Phys Rev. Lett. {\bf{97}}, 206805 (2006). 

\bibitem{Bradbury.prl.2006} F. R. Bradbury, A. M. Tyryshkin, G. Sabouret, J. Bokor, T. Schenkel, and S. A. Lyon, Phys Rev. Lett. {\bf{97}}, 176404 (2006). 

\bibitem{Schofield.prl.2003} S. R. Schofield, N. J. Curson, M. Y. Simmons, F. J. Ruess, T. Hallam, L. Oberbeck, and R. G. Clark, Phys. Rev. Lett. {\bf{91}}, 136104 (2003). 

\bibitem{Jamieson.apl.2005} D. N. Jamieson, C. Yang, T. Hopf, S. M. Hearne, C. I. Pakes, S. Prawer, M. Mitic, E. Gauja, S. E. Andresen, F. E. Hudson, A. S. Dzurak, and R. G. Clark, Appl. Phys. Lett. {\bf{86}}, 202101 (2005).

\bibitem{Koiller.prl.2002} B. Koiller, X. Hu, and S. Das Sarma, Phys Rev. Lett. {\bf{88}}, 027903 (2002).

\bibitem{Friesen.prl.2005} M. Friesen, Phys. Rev. Lett. {\bf{94}}, 186403 (2005).

\bibitem{Wellard.prb.2005} C. J. Wellard and L. C. L. Hollenberg, Phys. Rev. B {\bf{72}}, 085202 (2005).

\bibitem{Martins.prb.2004} A. S. Martins, R. B. Capaz, and Belita Koiller, Phys. Rev. B {\bf{69}}, 085320 (2004).

\bibitem{Rahman.prl.2007} R. Rahman, C. J. Wellard, F. R. Bradbury, M. Prada, J. H. Cole, G. Klimeck, and L. C. L. Hollenberg, Phys Rev. Lett. {\bf{99}}, 036403 (2007). 

\bibitem{Kane.nature.1998} B. E. Kane, Nature, {\bf{393}}, 133 (1998).

\bibitem{Vrijen.pra.2000} R. Vrijen, E. Yablonovitch, K. Wang, H. W. Jiang, A. Balandin, V. Roychowdhury, T. Mor, and D. DiVincenzo, Phys. Rev. A {\bf{62}}, 012306 (2000).

\bibitem{Skinner.prl.2003} A. J. Skinner, M. E. Davenport, and B. E. Kane, Phys. Rev. Lett. {\bf{90}}, 087901 (2003).

\bibitem{Hollenberg.prb.2004} L. C. L. Hollenberg, A. S. Dzurak, C. Wellard, A. R. Hamilton, D. J. Reilly, G. J. Milburn, and R. G. Clark, Phys. Rev. B {\bf{69}}, 113301 (2004). 

\bibitem{Andresen.Nanoletter.2007} S. E. S. Andresen, R. Brenner, C. J. Wellard, C. Yang, T. Hopf, C. C. Escott, R. G. Clark, A. S. Dzurak, D. N. Jamieson, and L. C. L. Hollenberg, Nano Lett. {\bf{7}}, 2000 (2007). 

\bibitem{Wellard.prb.2006} C. J. Wellard, L. C. L. Holllenberg, and S. Das Sarma, Phys. Rev. B {\bf{74}}, 075306 (2006). 

\bibitem{Greentree.prb.2004} A. D. Greentree, J. H. Cole, A. R. Hamiltom, and L. C. L. Hollenberg, Phys. Rev. B {\bf{70}}, 235317 (2004). 

\bibitem{Rahman.prb.2009} R. Rahman, S. H. Park, J. H. Cole, A. D. Greentree, R. P. Muller, G. Klimeck, and L. C. L. Hollenberg, Phys Rev. B {\bf{80}}, 035302 (2009).

\bibitem{Hollenberg.prb.2006} L. C. L. Hollenberg, A. D. Greentree, A. G. Fowler, and C. J. Wellard, Phys. Rev. B {\bf{74}}, 045311 (2006).

\bibitem{Xuedong.prb.2005} X. Hu, B. Koiller, and S. Das Sarma, Phys. Rev. B {\bf{71}}, 235332 (2005). 

\bibitem{Kohn.physrev.1955} W. Kohn and J.M. Luttinger, Phys. Rev. {\bf{98}}, 915 (1955).

\bibitem{Koiller.prb.2006} Belita Koiller, X. Hu, and S. Das Sarma, Phys. Rev. B {\bf{73}}, 045319 (2006). 

\bibitem{Ise.TCAD} Technology Computer Aided Design (TCAD), Integrated Systems Engineering AG, Zurich.

\bibitem{Wellard.prb.2003} C. J. Wellard, L. C. L. Hollenberg, F. Parisoli, L. M. Kettle, H.-S. Goan, J. A. L. McIntosh, and D. N. Jamieson, Phys. Rev. B {\bf{68}}, 195209 (2003). 

\bibitem{Slater.physrev.1954} J. C. Slater and G.F. Koster, Phys. Rev. Vol. {\bf{94}}, No. 6 (1954).  

\bibitem{Boykin.prb.2004} T. B. Boykin, G. Klimeck, and F. Oyafuso, Phys. Rev. B {\bf{69}}, 115201 (2004).

\bibitem{Klimeck.cmes.2002} G. Klimeck, F. Oyafuso, T. B. Boykin, R. C. Bowen, and P. von Allmen, CMES {\bf{3}}, 601, (2002). The TB calculations have been done using the NEMO 3D tool. 

\bibitem{Lee.prb.2004} S. Lee, F. Oyafuso, P. von Allmen, and G. Klimeck, Phys. Rev. B {\bf{69}}, 045316 (2004).

\bibitem{nanohub.note} nanoHUB.org computational resource of a 256-node 3.3GHz Pentium Irvindale PC cluster was used. The tight-binding calculations were done with the NEMO-3D tool.

\bibitem{Klimeck.ted.2007} G. Klimeck, S. Ahmed, N. Kharche, M. Korkusinski, M. Usman, M. Prada, and T. B. Boykin, IEEE Trans. Electron Dev. {\bf{54}}, 2079, (2007). 

\bibitem{Calderon.prl.2006} M. J. Calderon, B. Koliller, X. Hu, and S. Das Sarma, Phys. Rev. Lett. {\bf{96}}, 096802 (2006)

\bibitem{Kettle.prb.2003} L. M. Kettle, H. -S. Goan, S. C. Smith, C. J. Wellard, L. C. L. Hollenberg, and C. I. Pakes, Phys. Rev. B {\bf{68}}, 75317 (2003).  

\bibitem{Smit.prb.2004} G. D. J. Smit, S. Rogge, J. Caro, and T. M. Klapwijk, Phys. Rev. B {\bf{70}}, 035206 (2004).

\end{thebibliography}
\end{document}